\documentclass[sigconf,10pt]{acmart}

\AtBeginDocument{%
  }

\usepackage{hyperref}
\usepackage{multirow}
\usepackage{graphicx}

\usepackage{caption}
\usepackage{subcaption}

\usepackage[utf8]{inputenc}
\usepackage[ruled, lined, longend, linesnumbered]{algorithm2e}
\usepackage{amsmath,multicol}

\usepackage{amssymb,textcomp,comment}
\usepackage{makecell}
\usepackage{ulem}

\usepackage{booktabs}
\usepackage{multirow}
\usepackage{color}
\usepackage{colortbl}

\usepackage{listings}
\usepackage{enumitem}
\usepackage{cleveref}

\usepackage[ruled,linesnumbered]{algorithm2e}

\crefname{section}{§}{§§}
\Crefname{section}{§}{§§}

\definecolor{dkgreen}{rgb}{0,0.6,0}
\definecolor{gray}{rgb}{0.5,0.5,0.5}
\definecolor{mauve}{rgb}{0.58,0,0.82}

\usepackage{array}
\newcolumntype{L}[1]{>{\raggedright\let\newline\\\arraybackslash\hspace{0pt}}m{#1}}
\newcolumntype{C}[1]{>{\centering\let\newline\\\arraybackslash\hspace{0pt}}m{#1}}
\newcolumntype{R}[1]{>{\raggedleft\let\newline\\\arraybackslash\hspace{0pt}}m{#1}}

\newcommand{\sys}{\texttt{LLMCad}\xspace}
\newcommand{\largemodel}{target\xspace}
\newcommand{\smallmodel}{memory-resident\xspace}

\newcommand{\xdl}[1]{\textbf{\color{blue}{#1}}}
\newcommand{\mwx}[1]{\textbf{\color{red}{#1}}}
\newcommand{\shit}[1]{\textbf{\color{brown}{#1}}}

\newcommand{\todo}[1]{\textbf{\color{blue}{TODO: #1}}}

\setcopyright{none}
\renewcommand\footnotetextcopyrightpermission[1]{} 
\begin{document}

\title{\sys: Fast and Scalable On-device Large Language Model Inference}

\author{Daliang Xu$^{\blacklozenge}$, Wangsong Yin$^{\blacklozenge}$, Xin Jin$^{\blacklozenge}$, Ying Zhang$^{\blacklozenge}$, Shiyun Wei$^{\bigstar}$, Mengwei Xu$^{\Diamond}$, Xuanzhe Liu$^{\blacklozenge}$}
\affiliation {
	\institution{$^\blacklozenge$Key Lab of High Confidence Software Technologies (Peking University), Beijing, China}
	\country{}
}
\affiliation {
	\institution{$^\bigstar$Zhongguancun Laboratory, Beijing, China.}
	\country{}
}
\affiliation {
	\institution{$^\Diamond$State Key Laboratory of Networking and Switching Technology (BUPT), Beijing, China}
	\country{}
}

\email{{xudaliang, hg, xinjinpku, zhang.ying, weishiyun, liuxuanzhe}@pku.edu.cn}
\email{yws@stu.pku.edu.cn}
\email{
	mwx@bupt.edu.cn
}
\begin{abstract}

Generative tasks, such as text generation and question answering, hold a crucial position in the realm of mobile applications. Due to their sensitivity to privacy concerns, there is a growing demand for their execution directly on mobile devices. Currently, the execution of these generative tasks heavily depends on Large Language Models (LLMs). Nevertheless, the limited memory capacity of these devices presents a formidable challenge to the scalability of such models.

In our research, we introduce \sys, an innovative on-device inference engine specifically designed for efficient generative Natural Language Processing (NLP) tasks. The core idea behind \sys revolves around model collaboration: a compact LLM, residing in memory, takes charge of generating the most straightforward tokens, while a high-precision LLM steps in to validate these tokens and rectify any identified errors.
\sys incorporates three novel techniques:
(1) Instead of generating candidate tokens in a sequential manner, \sys employs the smaller LLM to construct a token tree, encompassing a wider range of plausible token pathways. Subsequently, the larger LLM can efficiently validate all of these pathways simultaneously.
(2) It employs a self-adjusting fallback strategy, swiftly initiating the verification process whenever the smaller LLM generates an erroneous token.
(3) To ensure a continuous flow of token generation, \sys speculatively generates tokens during the verification process by implementing a compute-IO pipeline.
Through an extensive series of experiments, \sys showcases an impressive token generation speed, achieving rates up to 9.3$\times$ faster than existing inference engines.

\end{abstract}
\maketitle

\section{introduction}

Generative tasks like text generation, question answering, and translation play a crucial role on mobile devices, as numerous applications rely on them to deliver key functionalities.
For instance, input method application like Google GBoard heavily leverages its text generation capabilities, while private assistant like Apple Siri uses it for question answering.
Such tasks are often privacy-sensitive and heavily rely on users' private data, thereby necessitating on-device local inference.

\begin{figure}
    \centering
    \begin{subfigure}[t]{0.48\textwidth}
    \includegraphics[width=\textwidth]{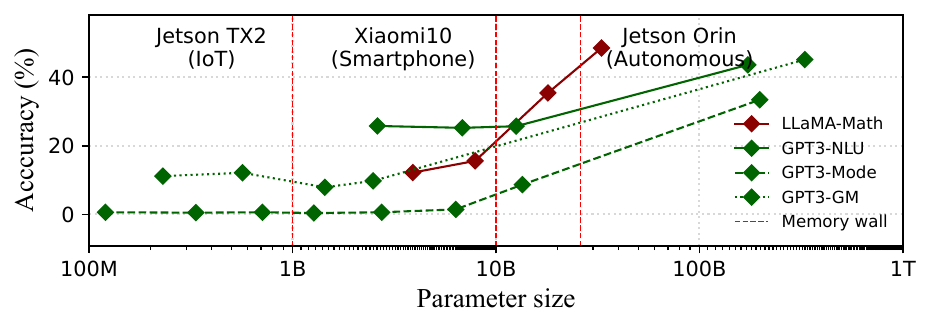}
    \vspace{-20pt}
    \caption{Emergent abilities across various LLMs.}
    \end{subfigure}
    \begin{subfigure}[t]{0.48\textwidth}
    \includegraphics[width=\textwidth]{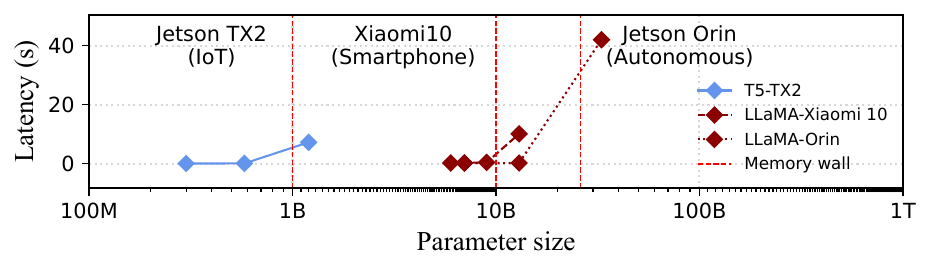}
    \vspace{-20pt}
    \caption{Generation latency of LLMs across various devices.}
    \end{subfigure} 
    \vspace{-10pt}
    \caption{The memory wall hinders LLM's ``scaling law'' on mobile devices.
*-Math, *-NLU, *-Mode, and *-GM denote LLMs' emergent abilities: math reasoning, multi-task comprehension, mode arithmetic, and learning meaningful representations. }
    \vspace{-10pt}
    \label{fig-intro}
\end{figure}

Large language models (LLMs), especially those built atop transformer decoder~\cite{vaswani2017attention} such as GPT-3~\cite{brown2020language} and LLaMA~\cite{touvron2023llama}, have become the de-facto approach to solve NLP generative tasks. 
Recent research in the machine learning community has demonstrated that scaling up such LLMs parameter size brings accuracy improvement and emergent ability~\cite{wei2022emergent,wei2022emergent,touvron2023llama,brown2020language,wei2022chain}, as shown in Figure~\ref{fig-intro}(a).
In general, an LLM necessitates more than 1B parameters to learn meaningful representations~\cite{patel2021mapping},  over 10B parameters to exhibit certain arithmetic reasoning abilities~\cite{cobbe2021training}, and more than 30B parameters to achieve multi-task comprehension capabilities~\cite{hendrycks2020measuring}.
This phenomenon is well-recognized in the machine learning community as the \textit{scaling law}~\cite{kaplan2020scaling,aghajanyan2023scaling,clark2022unified,alabdulmohsin2022revisiting}.

\textbf{Key challenge: memory wall.}
However, our preliminary experiments in Figure~\ref{fig-intro}(b) reveal that the scaling ability is challenged on mobile devices.
Specifically, when LLMs are too large to be fit into device memory, mobile DNN engines like MNN~\cite{mnn}a nd llama.cpp~\cite{llama-cpp} need to repetitively release and load model weights.
It results in 59--224$\times$ lengthened inference latency.
Such memory wall severely hinders the scaling law.
Users have to choose between real-time generation and emergent ability. 
For instance, 10B parameters represent the minimum size required for LLaMA to possess arithmetic reasoning capabilities, yet it also represents the maximum parameter size for achieving real-time inference on smartphones (e.g., Xiaomi 10).

\textbf{\sys: breaking memory wall through model collaboration.}
In this paper, we propose \sys, the first efficient inference engine for on-device generative NLP tasks.
\sys delivers LLM's scaling ability to mobile devices with a tolerable generation speed through \textit{model collaboration}.
The main idea is to delegate most tokens to a smaller real-time LLM that can be totally hosted in device memory (namely \smallmodel LLM).
The design is based on a key observation that, while a smaller LLM is inadequate to deliver satisfactory end-to-end sentences, they can correctly generate most easy tokens (e.g., determiners, pronouns, and punctuations).
Furthermore,
LLMs are often trained with a series of model variants, e.g. T5-Small/Base/Large~\cite{raffel2020exploring} and LLaMa-7B/13B/33B~\cite{touvron2023llama}, and its smaller counterpart (e.g., LLaMa-7B and T5-small, dubbed \textit{\smallmodel model} in this paper) can often be hosted in memory easily~\cite{touvron2023llama,xue2020mt5,raffel2020exploring,brown2020language}.


\sys employs a unique form of model collaboration, namely ``generate-then-verify''~\cite{leviathan2023fast,chen2023accelerating}.
In this approach, the \smallmodel LLM serves as a token generator, while a \largemodel LLM acts as a verifier, using its output as the ground truth to inspect and rectify any errors introduced during the token generation process.
This approach provides two significant advantages:
(1) \textit{No compromising accuracy.}
Each token is verified by the \largemodel model, therefore its accuracy is guaranteed.
This is crucial as a wrong token could propagate its error to the subquent tokens due to the autoregressive nature.
(2) \textit{Fast verification.} As will be detailed in $\S$\ref{sec-bgd-chanllenge}, the verification of $N$ tokens can be accomplished within one-shot inference o f the \largemodel model, therefore much faster than using it to generate $N$ tokens sequentially.

Despite these advantages, applying model collaboration for on-device LLM introduces three distinctive challenges:

\begin{itemize} [leftmargin=*,topsep=2pt]
    \item 
    \textit{Overlooked correct tokens with sub-optimal confidence.}
    Typically, state-of-the-art LLM engines and studies always use the token with the highest probability as the output. 
    Nevertheless, our observation has revealed that some of generation errors by the \smallmodel LLM can be rectified by the sub-optimal tokens.
    Figure~\ref{fig-translation-case} gives a real-world example of such phenomenon.
    Given the significant performance overhead associated with on-device verification, \sys must capitalize on these often-overlooked tokens to reduce the frequency of verification. 
    
    \item 
    \textit{Verification timing.}
    Another crucial aspect is determining when to initiate the verification process.
    On-device verification is time-consuming, e.g., taking 7.1s on Jetson TX2. 
    Too early or too late verification just wastes computing mobile devices scarce resources by invalid verification (i.e., no errors detected) or useless tokens. 
    Prior works have typically relied  either a single token or token sequence length, which may not accurately pinpoint the optimal verification timing.

    \item \textit{IO vs. compute asymmetry.}
    With a LLM cascade, the large LLM execution blocks the small model inference due to the cross-token dependency, and the processor is under-utilized as the I/O bottlenecks during weights loading.
    Such a situation severely hampers the inference speed as the \largemodel model needs to be invoked unavoidably to guarantee correct tokens generation.
\end{itemize}

In response, \sys desgins three novel techniques:

\noindent \textbf{(1) Token tree generation and verification ($\S$\ref{sec-design-tree-generation-verification}).}
Instead of generating and verifying a linear token sequence, \sys employs a different approach by constructing and validating a ``token tree.'' 
This token tree permits each token to have multiple potential succeeding tokens.
To accomplish this efficiently, \sys employs three novel modules:
(1) \texttt{Confidence-based branch pacer} paces the progress of different branches to prevent the wasteful allocation of computing resources to the wrong branch;
(2) \texttt{Tree decoder} generates tokens from various branches without incurring the overhead of context switching between them;
(3) \texttt{Non -autoregressive token tree verifier} examines and rectifies all errors within a token tree in a batch manner, at the cost of a single iteration.

\noindent \textbf{(2) Self-adaptive fallback strategy ($\S$\ref{sec-design-fallback-strategy}).} 
This strategy is devised to initiate the verification process promptly when the \smallmodel LLM generates an incorrect token.
It is inspired by two key observations:
(1) Typically, each token generated by the \smallmodel LLM introduces some ``uncertainty'' (imperfect confidence score).
\sys uses a more accurate metric referred to as \textit{cumulative uncertainty} within the token tree compared.
Compared to prior works, this metric better reflects the error probability associated with \smallmodel LLM generation, especially considering the accumulative nature of autoregressive models.
(2) Historical data pertaining to the accuracy of verified tokens is harnessed to assess the \smallmodel LLM's generation capability. 
A stronger generation ability necessitates a lower frequency of verification.

\noindent \textbf{(3) Speculative generation pipeline ($\S$\ref{sec-design-speculative}).}
To break the cross-token dependency and enhance parallelism, we propose \textit{speculative generation}, i.e., continuing generating tokens through the \smallmodel LLM during the verification process, 
This is founded on the insight that sometimes the verification process may not detect errors, rendering the speculatively generated tokens usable. 
However, simultaneous speculative generation with verification directly can lead to processor and memory contentions.
To further tackle this issue, \sys incorporates a fine-grained pipeline, ensuring that the speculative generation only runs when loading \largemodel LLM parameters below the memory upper bound to void interfering with the regular verification process.

\textbf{Implementation and evaluation.} We have fully implemented \sys on top of two SOTA LLM engines: PyTorch~\cite{pytorch} and  llama.cpp~\cite{llama-cpp}.
Extensive evaluation of the system was conducted across four platforms: two IoT devices (Jetson TX2 and Jetson Orin NX) and two smartphones (Xiaomi 10 and Xiaomi 11).
This evaluation encompassed six widely utilized LLMs (GPT2~\cite{radford2019language}, T5~\cite{raffel2020exploring}, mT5~\cite{xue2020mt5}, Bart~\cite{DBLP:journals/corr/abs-1910-13461}, Vicuna, and LLaMa2~\cite{touvron2023llama}) and seven datasets (CNN/Daily~\cite{see-etal-2017-get}, Wikitext~\cite{merity2016pointer}, iwlt2017~\cite{cettolo-etal-2017-overview}, wmt14/22~\cite{bojar-EtAl:2014:W14-33}, SQuAD~\cite{2016arXiv160605250R}, parrot, and TruthfulQA~\cite{lin2021truthfulqa}).
We also compared \sys with five state-of-the-art competitive baselines~\cite{pytorch,llama-cpp,leviathan2023fast,kim2023big,guo2023sti}, encompassing two computing-loading pipeline frameworks and two  ``generator and verifier'' LLM collaboration frameworks.
Our results unequivocally demonstrate \sys's superior performance.
When compared to the state-of-the-art LLM engines, \sys can reduce average per-token generation time by 2.9--9.3$\times$ and 3.5--4.7$\times$ on IoT devices and smartphones, respectively, without comprising accuracy.
For $>$10B-sized LLMs like LLaMA2-13B that are previously unbearable on smartphones, \sys generates more than one token per second.
Furthermore, compared with competitive baselines, \sys can achieve 
up to 5.6$\times$ speedup and noticeable higher accuracy.

The major contributions of this work are as follows:
\begin{itemize}[leftmargin=10pt,topsep=0pt]
\item We thoroughly explore the opportunities and challenges of inferring LLMs on the device. 
\item We propose \sys, the first efficient inference engine for on-device generative NLP tasks.
To speedup the generation procedure, \sys uses ``generator and verifier'' LLM collaboration  and incorporates three novel techniques: tree generation and verification, self-adaptive fallback strategy, and speculative generation pipeline.
These advancements enable \sys to effectively mitigate the memory wall problem.
\item We prototype \sys and evaluate it with representative LLMs and commodity mobile devices. 
The results demonstrate its superior performance over existing methods.
\end{itemize}

\section{background and motivation}

\subsection{Decoder-based Generative LLMs} \label{sec-bgd-generative}




\noindent \textbf{On-device generative ML tasks.}
Generative tasks are those involving automatic generation or synthesis of new contents like text sequences and image pixels~\cite{rombach2022high,reed2016generative,bau2020semantic}. 
Typical generative tasks in the NLP domain (focus of this work) include language modeling, machine translation, summarization, and question answering. 
The key focus is to provide novel, meaningful, and coherent content output.
As compared to traditional classification tasks such as topic classification and image classification, generative tasks are often more challenging but also more profound to human lives~\cite{factGenerativeAdvantages}. 

Generative ML tasks have been widely served for mobile users, such as language modeling for Google Gboard~\cite{googleGboardGoogle}, question answering for Siri~\cite{appleSiri}, and translation services like iTranslate~\cite{itranslateLeadingTranslation} and Google Translate~\cite{googleGoogleTranslate}, etc.
To guarantee user data privacy (e.g., text corpus) and service availability, the models are better to be deployed on devices directly for local inference.

\begin{figure}[t]
	\centering
        \begin{subfigure}[T]{0.17\textwidth}
        \includegraphics[width=\linewidth]{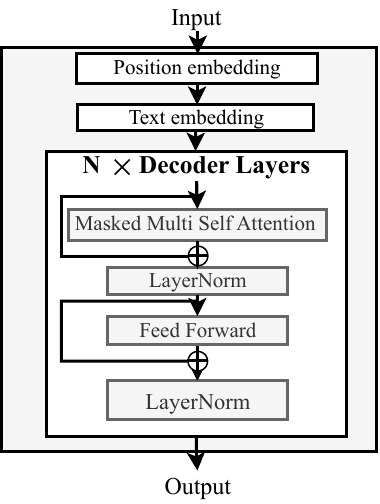}
        \caption{GPT3 architecture}         
        \end{subfigure}
	\begin{subfigure}[T]{0.275\textwidth}
    \includegraphics[width=\textwidth]{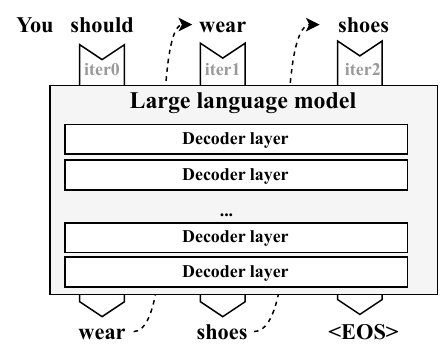} 
    \caption{Autoregressive inference}
    \end{subfigure}
    \vspace{-10pt}
	\caption{The architecture of decoder-only language model (GPT3) and the overview of LLM inference pattern: autoregressive.}
    \vspace{-10pt}
	\label{fig-gpt3-architecture}
\end{figure}
\noindent \textbf{Decoder-based LLM architecture.}
Decoder-based large language model (LLM), including both decoder-only and encoder-decoder architectures, is the de-facto approach for generative tasks, such as GPT-3~\cite{brown2020language}, LLaMa~\cite{touvron2023llama}, and GLM-130B~\cite{zeng2022glm,du2022glm}.
As shown in Figure~\ref{fig-gpt3-architecture}(a),
a typical decoder-based LLM consists of a text embedding, a position embedding, and many sequentially stacked decoder layers, where each decoder layer includes masked self-attention, LayerNorm, and Linear operations.
For those encoder-decoder LLMs such as T5~\cite{raffel2020exploring} and mT5~\cite{xue2020mt5}, encoder layers are incorporated before the decoder to enhance semantic understanding capabilities.

\noindent \textbf{Autoregressive inference.}
Generative LLMs employ an \textit{autoregressive} inference procedure that generates one token at a time and takes that token as input to generate the next one.
For instance, Figure~\ref{fig-gpt3-architecture}(b) illustrates a three-autoregressive-iteration inference procedure.
In the 1st iteration, the model takes all existing tokens  ("You should") as input and generates the output "wear."
In the next iteration, the newly generated "wear" will be fed into the model, which then predicts "shoes."
This process continues until the model generates the end-of-sequence token ($<EOS>$), indicating the end of the generation procedure.
The nature of autoregressive inference introduces unique challenges for optimizing on-device LLM as will be described later.

\subsection{On-device LLM is Memory-bounded} \label{sec-bgd-memory}



In this section, we perform pilot experiments to reveal the performance issue of on-device LLM inference.
The experiments are performed on typical LLMs (GPT2, T5, and LLaMa), datasets (SQuAD and TriviaQA), and mobile devices (Jetson TX2 and Xiaomi 10) using state-of-the-art DL engines (PyTorch~\cite{pytorch} and llama.cpp~\cite{llama-cpp}.
We summarize our key findings below.

\textbf{Scaling up parameter size brings accuracy improvement.}
Transformer-based LLM architecture is highly flexible and scalable by simply adjusting the encoder/decoder layers, sequence length, and other hyper-parameters.
Consequently, popular LLM is often developed with a series of model variants, such as T5-Small/Base/Large~\cite{raffel2020exploring} and LLaMa-7B/13B/33B/65B~\cite{touvron2023llama}.
With the parameter size scaling up, the model exhibits stronger abilities.
As shown in Table~\ref{table-model-size}, T5-Large outperformed T5-Small by a significant margin, achieving a 7.6\% improvement in accuracy on the SQuAD dataset. 
Similarly, LLaMa-13B demonstrated a 6.6\% higher QA accuracy on TriviaQA than LLaMa-7B.
Indeed, such a phenomenon is well known in ML community as \textit{scaling law}~\cite{kaplan2020scaling,aghajanyan2023scaling,clark2022unified,alabdulmohsin2022revisiting}.

\begin{figure}
    \centering
    \includegraphics[width=0.48\textwidth]{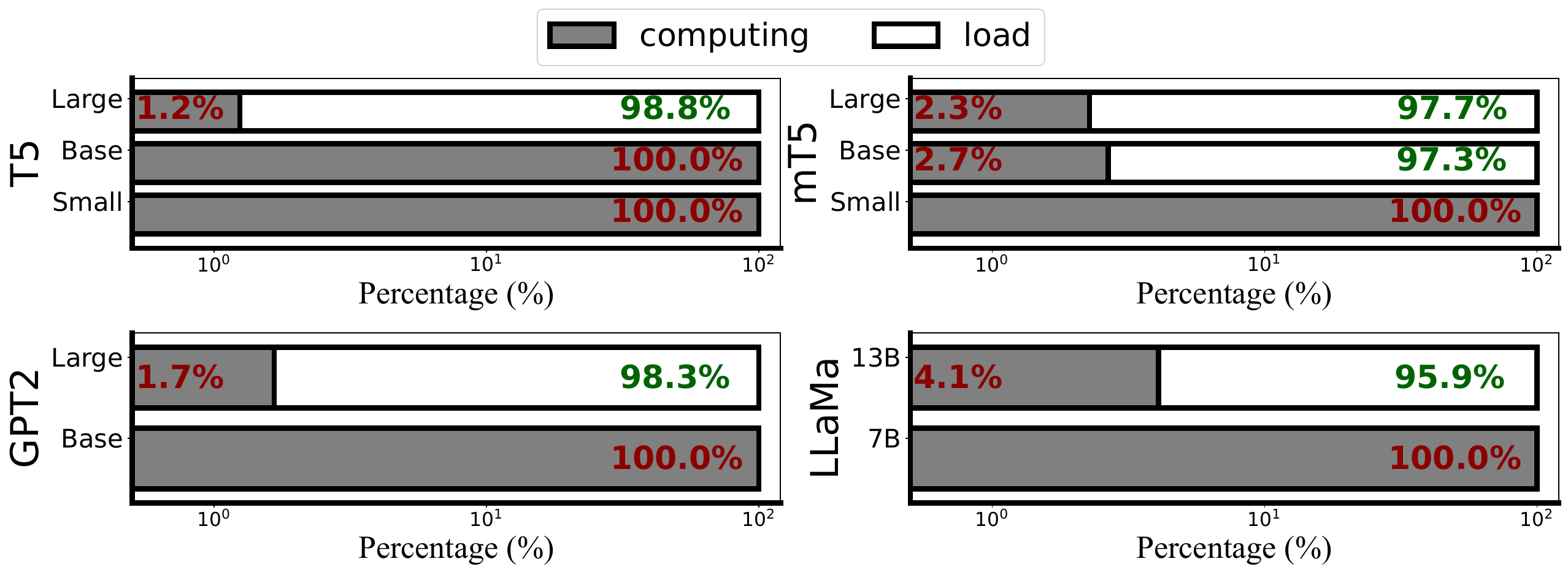}
    \vspace{-10pt}
    \caption{Inference delay breakdown of different LLM variants in one autoregressive iteration. }
    \vspace{-10pt}
    \label{fig-inference-breakdown}
\end{figure}

\textbf{On-device LLM scalability hinders on the memory wall.}
However, as shown in Table~\ref{table-model-size}, the inference speed declines rapidly when the memory consumption exceeds the memory budget.
For instance, on the TX2 device, the inference latency increases by 189--224$\times$ with only 5.8--12.2$\times$ increase in model size.
To gain a deeper understanding of the factors influencing inference speed, we conducted a breakdown analysis, as shown in Figure~\ref{fig-inference-breakdown}.
It clearly shows that, when the model inference demands a memory size unaffordable on edge devices, loading parameters from disk to memory (i.e., disk I/O) soon becomes the bottleneck of the inference time (95.9--98.8\%).
This situation attributes to the fact that the state-of-the-art device-side LLM engines, such as MNN~\cite{mnn} and llama.cpp~\cite{llama-cpp}, resort to the \textit{swapping} technique which dynamically releases the inferred weights memory and loads weights to be inferred  from the disk when memory constraints are exceeded.

\begin{table}[t]
\centering
\resizebox{0.48\textwidth}{!}{%
\begin{tabular}{lrrrr}
\toprule[0pt]
\multicolumn{5}{c}{\textbf{(a) On TriviaQA with Xiaomi10}}                                                                                                                                                   \\ \toprule[1.5pt]
\multicolumn{1}{l}{\textbf{}}  & \multicolumn{1}{r}{\textbf{\# of Params (B)}} & \multicolumn{1}{r}{\textbf{Accuracy}} & \multicolumn{1}{r}{\textbf{PeakMem. (GB)}} & \textbf{Infer. Time (ms)} \\ \hline
\multicolumn{1}{l}{
\small{LLaMa-7B (4bits)}}   & \multicolumn{1}{r}{7}                         & \multicolumn{1}{r}{50.0}              & \multicolumn{1}{r}{4.1}                    & 275 (1x)                  \\ 
\multicolumn{1}{l}{\small{LLaMa-13B (4bits)}}  & \multicolumn{1}{r}{13}                        & \multicolumn{1}{r}{56.6}              & \multicolumn{1}{r}{9.8}                    & 10118 (37x)               \\ 
\multicolumn{1}{l}{\small{LLaMa-33B (4bits)}}  & \multicolumn{1}{r}{33}                        & \multicolumn{1}{r}{65.1}              & \multicolumn{1}{r}{20.8}                   &   22017 (87x)                 \\ \toprule[1.5pt]
& \multicolumn{1}{l}{}                           & \multicolumn{1}{l}{}                   & \multicolumn{1}{l}{}                        & \multicolumn{1}{l}{}      \\ \toprule[0pt]
\multicolumn{5}{c}{\textbf{(b) On SQuAD with TX2}}                                                                                                                                                           \\ \toprule[1.5pt]
\multicolumn{1}{l}{\textbf{}}  & \multicolumn{1}{r}{\textbf{\# of Params (B)}} & \multicolumn{1}{r}{\textbf{Accuracy}} & \multicolumn{1}{r}{\textbf{PeakMem. (GB)}} & \textbf{Infer. Time (ms)} \\  \hline
\multicolumn{1}{l}{GPT2}       & \multicolumn{1}{r}{0.14}                      & \multicolumn{1}{r}{49.8}              & \multicolumn{1}{r}{0.5}                    & 37.64 (1x)                \\ 
\multicolumn{1}{l}{GPT2-Large} & \multicolumn{1}{r}{0.8}                       & \multicolumn{1}{r}{53.7}              & \multicolumn{1}{r}{3.1}                    & 8065 (214x)               \\ \toprule[1.5pt]
                                & \multicolumn{1}{l}{}                           & \multicolumn{1}{l}{}                   & \multicolumn{1}{l}{}                        & \multicolumn{1}{l}{}      \\ \toprule[1.5pt]
\multicolumn{1}{l}{mT5-Small}  & \multicolumn{1}{r}{0.3}                       & \multicolumn{1}{r}{76.4}              & \multicolumn{1}{r}{0.7}                    & 31 (1x)                   \\ 
\multicolumn{1}{l}{mT5-Base}   & \multicolumn{1}{r}{0.58}                      & \multicolumn{1}{r}{83.8}              & \multicolumn{1}{r}{1.5}                    & 2134 (69x)                \\ 
\multicolumn{1}{l}{mT5-Large}  & \multicolumn{1}{r}{1.2}                       & \multicolumn{1}{r}{87.0}              & \multicolumn{1}{r}{3.9}                    & 7214 (230x)               \\ \toprule[1.5pt]
                                & \multicolumn{1}{l}{}                           & \multicolumn{1}{l}{}                   & \multicolumn{1}{l}{}                        & \multicolumn{1}{l}{}      \\ \toprule[1.5pt]
\multicolumn{1}{l}{T5-Small}   & \multicolumn{1}{r}{0.06}                      & \multicolumn{1}{r}{79.1}              & \multicolumn{1}{r}{0.2}                    & 27 (1x)                   \\ 
\multicolumn{1}{l}{T5-Base}    & \multicolumn{1}{r}{0.22}                      & \multicolumn{1}{r}{85.4}              & \multicolumn{1}{r}{0.8}                    & 37 (1.4x)                 \\ 
\multicolumn{1}{l}{T5-Large}   & \multicolumn{1}{r}{0.73}                      & \multicolumn{1}{r}{86.7}              & \multicolumn{1}{r}{2.8}                    & 7098 (263x)               \\ \toprule[1.5pt]
\end{tabular}%
}
\caption{The parameter size, accuracy, peak memory usage, and inference latency of LLM variants in one autoregressive iteration. (a) experiments are performed on PyTorch while (b) experiments are performed on llama.cpp. }
\vspace{-20pt}
\label{table-model-size}
\end{table}

\textbf{The autoregressive nature makes traditional memory optimizations barely effective for generative LLM.}
It is worth noting that memory optimization for model inference has been a well-researched topic in recent years~\cite{yu2022orca,huang2020swapadvisor,meng2017training,guo2023sti}. 
Various system-level methods have been explored, such as batching~\cite{yu2022orca}, compute-I/O pipeline~\cite{guo2023sti}, and smart swapping~\cite{huang2020swapadvisor,meng2017training}. 
However, these works can hardly apply to on-device LLM, because:
(1) Parallelization/batching is not available as autoregressive inference requires generating tokens sequentially;
(2) Overlapping can gain limited benefits since I/O time is over a hundredfold larger than computing.
Additionally, algorithmic approaches like quantization~\cite{wang2021spatten,frantar2022gptq,yao2022zeroquant,hu2021lora} can bring a few times memory reduction (e.g., FP16$\xrightarrow{}$INT4~\cite{frantar2022gptq}), their efficiency is limited as low-bit precision (e.g., 2 bits) has been demonstrated to be insufficient to retain model capability~\cite{frantar2022gptq,yao2022zeroquant,kim2023full}.
Note that the proposal of this work is at system level and is compatible with quantization.

\begin{figure*}
    \centering
    \includegraphics[width=0.98\textwidth]{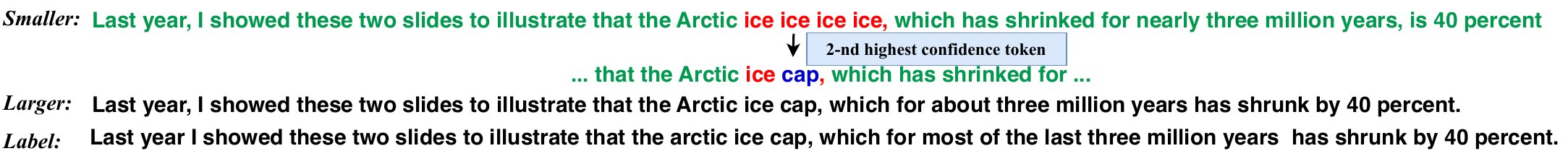}
    \vspace{-5pt}
    \caption{A translation generation example from mt5-small and mt5-large models as well as its ground truth label. Green:  correct parts of small model generation; Red: error propagation parts of small model generation; Blue: the sub-optimal token which is the correct answer.
    Noticeably, on the iwslt2017 de-en translation dataset~\cite{cettolo-etal-2017-overview}, the mT5-small model correctly generates nearly 69.3\% of tokens, while the number for the mT5-large model is 73.1\%.}
    \label{fig-translation-case}
\end{figure*}

\subsection{Opportunities and Challenges of LLM collaboration} \label{sec-bgd-chanllenge}

This work focuses on \textit{model collaboration} approach~\cite{schuster2022confident,wang2023tabi,lee2019mobisr}, which leverages multiple models with accuracy-cost tradeoff to speed up inference.
In the case of generative NLP tasks, we delegate most computations (i.e., tokens) to a smaller model that entirely fits into the memory budget.
The key rationale is that smaller models can exhibit close performance to the large one, especially for the easier data points~\cite{han2021legodnn,fang2018nestdnn}.
Our empirical experiments have confirmed such an assumption:
in the iwslt2017 de-en translation dataset~\cite{cettolo-etal-2017-overview}, mT5-Small correctly generates more than 80\% tokens as mT5-Large does.
Figure~\ref{fig-translation-case} gives one concrete example of translating a sentence by the smaller model and larger LLM, and the corresponding ground truth.
It shows that most of the tokens (in green) generated by the small model are correct.

However, employing model collaboration for LLMs faces one critical challenge: Wrong token delegation could be fatal.
Traditional model cascade either relies on internal~\cite{schuster2022confident} or external knowledge~\cite{lee2019mobisr,wang2023tabi} to select a portion of data (often the easier ones) to be processed by the smaller model.
In such circumstance, the accuracy cannot be guaranteed.
For generative NLP tasks, however, a wrongly generated token by the small model could propagate the error to the subsequent ones and finally cause catastrophic results due to its autoregressive nature~\cite{vaswani2017attention,raffel2020exploring,radford2018improving}.
For example, in Figure~\ref{fig-translation-case}, the second "ice" in red is incorrect, resulting in additional two "ice" generations and wrong translation information in the subsequent tokens.
Note that generative NLP tasks are often accuracy-sensitive, e.g., translation and Q/A, as a wrongly generated result could misinform users and result in unexpected behaviors.

To tackle this issue, \sys employs a unique form of model collaboration, namely ``generate-then-verify''~\cite{leviathan2023fast,chen2023accelerating}.
In this approach, the \smallmodel LLM serves as a token generator, while a \largemodel LLM acts as a verifier, using its output as the ground truth to inspect and rectify any errors introduced during the token generation process.
By doing that, \sys can prevent from propagating its error to the subquent tokens due to the autoregressive nature.and ensure no compromising accuracy.

\section{Design}

\subsection{Overview} \label{sec-design-overview}

\sys is built on two LLMs:
a \largemodel model that is accurate but heavy (cannot fit to device memory) like mT5-large;
and a \smallmodel model that is less accurate but lightweight like mT5-small.
The design goal of \sys is to generate texts with the speed of \smallmodel model \textit{without compromising the accuracy} of \largemodel (larger) model.

\begin{figure}[t]
	\centering
        \includegraphics[width=0.48\textwidth]{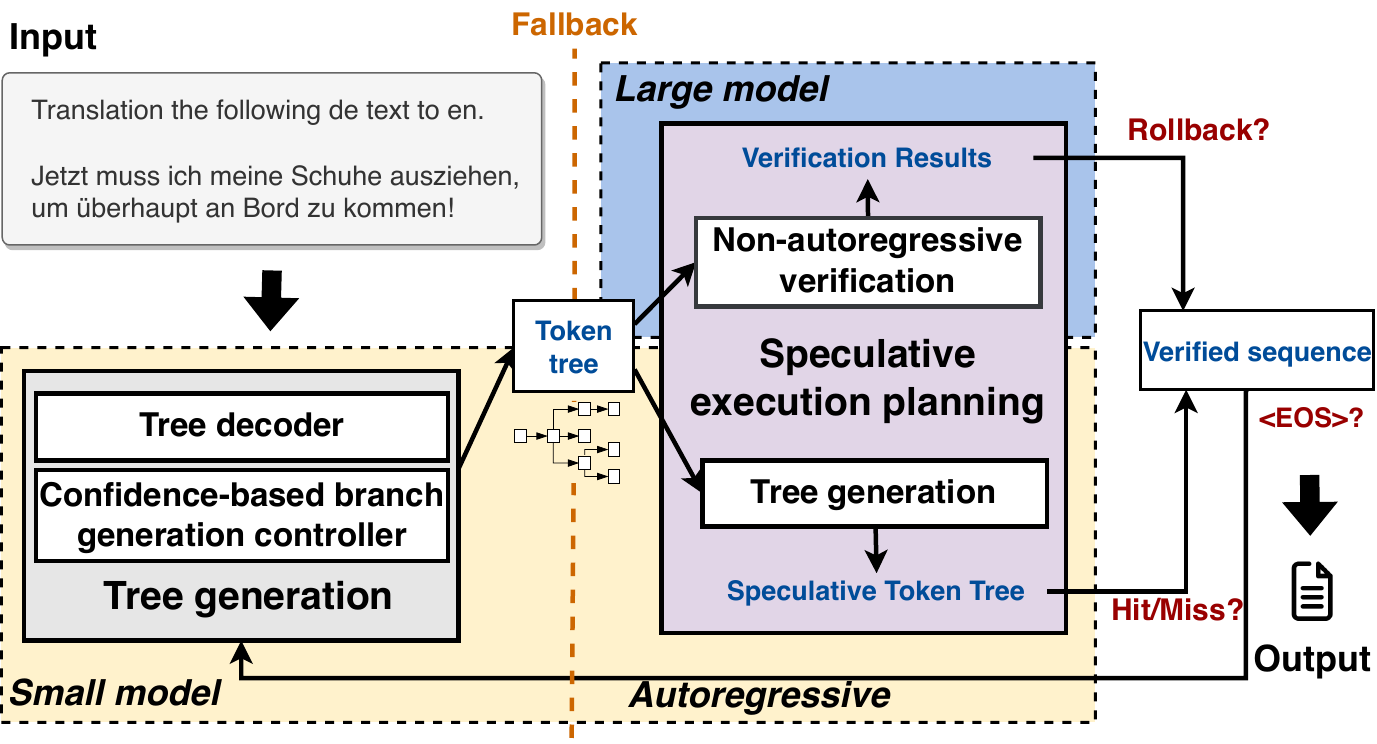}
    \vspace{-20pt}
	\caption{The workflow of \sys.}
     \vspace{-20pt}
	\label{fig-workflow}
\end{figure}
\begin{figure*}
    \centering
    \includegraphics[width=0.98\textwidth]{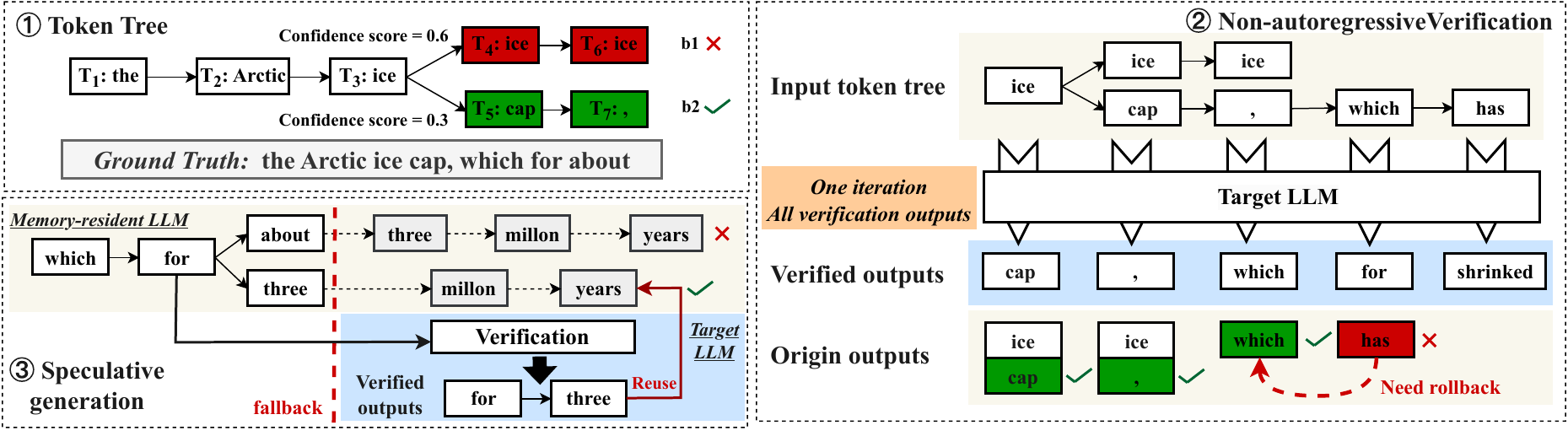}
    \vspace{-10pt}
    \caption{An illustrative example of \sys.}
    \label{fig-motivate-example}
\end{figure*}

\noindent \textbf{Simplified workflow and an illustrative example}
Figure~\ref{fig-workflow} illustrates the workflow of \sys.
Figure~\ref{fig-motivate-example} also provides an illustrative example based on the case of Figure~\ref{fig-translation-case} to exemplify the workflow. 
Essentially, \sys is a generation and verification framework using the \smallmodel LLM as a generator, and the \largemodel LLM as a  verifier.



First, \sys feeds the input text to the \smallmodel model and generates a \textit{token tree}. 
A \textit{token tree} is the intermediate result generated by the \smallmodel model (details in \textit{tree generation} $\S$\ref{sec-design-tree-generation-verification}).
Unlike a token sequence where each token has only a single succeeding token, a token in a token tree can have multiple candidate succeeding tokens, as shown in Figure~\ref{fig-motivate-example}\textcircled{1}.
Each of the candidate tokens represents a candidate token sequence (referred to as a \textit{branch}).
This is based on the observation that sometimes the ``sub-optimal'' tokens generated by the \smallmodel LLM is the true output of the \largemodel LLM, e.g., the alternative token ``cap''.
In practice, any candidate token with a confidence higher than a threshold (e.g., 30\%) generates a branch.


Each token generated by the \smallmodel LLM introduces some ``uncertainy'' (imperfect confidence score).
Once such uncertainty accumulates to a level in the output sentence, the \largemodel LLM is used to verify all the generated branches since last verification, as shown in Figure~\ref{fig-motivate-example}\textcircled{2}.
Notably, the verification of $N$ tokens can be done within one-shot inference with \largemodel LLM, therefore much faster than using it to generate one token by $N$ times.
such verification process is therefore termed ``non-autoregressive''.
Once an error is detected, \sys will rollback the token tree and rectify the it.
The details of the verification and rollback strategy is discussed in $\S$\ref{sec-design-tree-generation-verification}.


The verification process involves \largemodel LLM inference, therefore is I/O-bound as previously shown in $\S$\ref{sec-bgd-chanllenge}.
\sys further proposes \textit{Speculative generation} to exploit the under-utilized hardware resources by generating tokens \textit{speculatively} ($\S$\ref{sec-design-speculative}), i.e., continuing generating tokens through the \smallmodel model during verification process, as shown in Figure~\ref{fig-motivate-example}\textcircled{3} dashed boxes on the left side of the red dash line.
This approach is based on the insight that sometimes the verification detects no error so the speculatively generated tokens could be used afterwards.
Therefore, it can effectively hide execution latency under the I/O, e.g., the second branch in Figure~\ref{fig-motivate-example}\textcircled{3}.

\sys repeat the above generation and verification process until it encounters ``<EOS>'', the end token.

\begin{figure*}
    \centering
    \begin{subfigure}[t]{0.35\textwidth}
        \includegraphics[width=\textwidth]{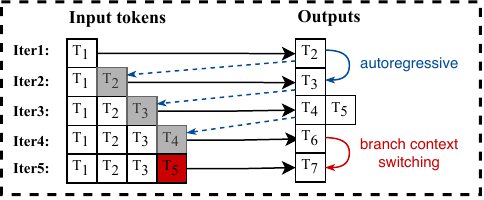}
        \caption{Existing tree generation procedure}
    \end{subfigure}
    \begin{subfigure}[t]{0.62\textwidth}
    \includegraphics[width=\textwidth]{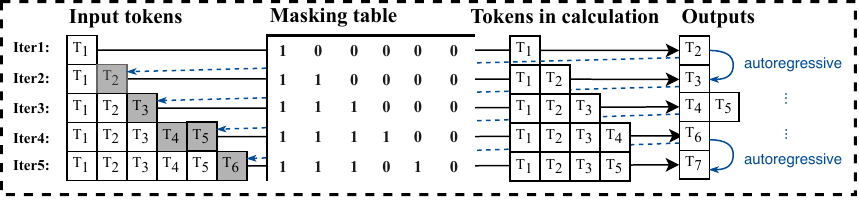}
    \caption{Tree decoder procedure}
    \end{subfigure}
    \vspace{-5pt}
    \caption{Examples of tree generation procedure in state-of-the-art LLM engines and the tree decoder procedure of \sys based on the case of Figure~\ref{fig-motivate-example}\textcircled{1}.}
    \vspace{-10pt}
    \label{fig-tree-masking}
\end{figure*}

\subsection{Token tree generation and verification} \label{sec-design-tree-generation-verification}


This subsection is mainly to discuss how the token tree is generated and verified in \sys.

\noindent \textbf{Token tree generation}
To generate useful token trees, \sys needs to answer two crucial questions:

\noindent $\bullet$ Branches compete for computing resource (e.g., GPU) to generate subsequent tokens by running \smallmodel model.
At a timestamp, which branch shall receive the resource to lengthen its token sequence?
The decision is crucial as generating tokens to a wrong branch (as verified by the \largemodel model later) wastes computing resource and delays the generation of true tokens.

\noindent $\bullet$ Generating tokens from different branches requires switching between branch contexts.
How to generate token from different branches efficiently?
The design is crucial as \sys needs to frequently switch between up to tens of branches.



In response, \sys incorporates two novel techniques:

\noindent $\bullet$ \textit{Confidence-based branch pacer.}
To properly pace the progress of different branches, \sys relies on the fact that branch with a higher probability is more likely to be the correct result and should have a longer sequence length.
Here, \sys models the probability with the cumulative confidence scores given by the \smallmodel for each token generated.
To control the branch length dynamically, \sys uses max-min fairness~\cite{hahne1991round}. 

Assuming that there are $N$ branches and $M$ tokens, and the i-th branch includes $T^B_i$ tokens, i-th branch cumulative confidence $C_i$ is the product of every token's confidence.
Thus, the branch length problem can be done by solving the following problem.
\begin{align}
    f(x) &= M * \frac{C_x}{\sum_{i=0}^N C_i } - T^B_x \\
    Obj &= min_{x=0}^N f(x)
\end{align}
Under the max-min fairness, \sys tries to allocate more hardware resources to the branch which is more likely to be the ground truth.



\begin{figure*}[t]
    \centering
    \includegraphics[width=0.96\textwidth]{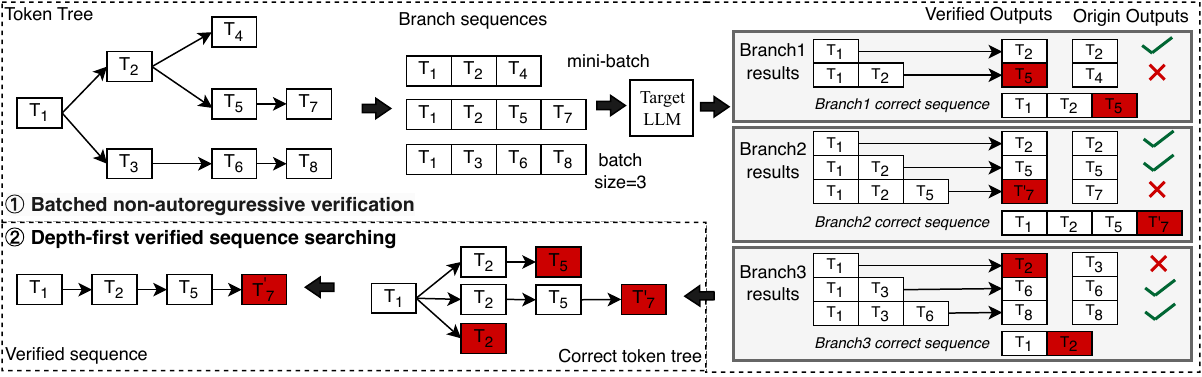}
    \vspace{-10pt}
    \caption{The illustration of token tree verification.}
    \vspace{-10pt}
    \label{fig-post-processing}
\end{figure*}

\noindent $\bullet$ \textit{Tree decoder.}
We commence our study by conducting an exhaustive analysis of the fundamental reasons behind the substantial performance overhead incurred by branch context switching (e.g., 25\% overhead for mT5 models on Jetson TX2).
In Figure~\ref{fig-tree-masking}(a), we provide an illustration of the implementation of branch context switching within state-of-the-art LLM engines, such as PyTorch, using the scenario depicted in Figure~\ref{fig-motivate-example}\textcircled{1} as a case study.
In this illustration, iterations 1--4 take the previous output token as the new input.
However, generating token ``7'' in iteration 5 necessitates branch switch from b1 to b2, which involves the removal of token $T_4$, the omission of the new token $T_6$, and the utilization of the sub-optimal output $T_5$ from iteration 3 as input. 
Consequently, \sys must deviate from the autoregressive rule and modify each iteration input with a substantial amount of metadata (e.g., Key-Value cache~\cite{yu2022orca,kim2023full,vllm} and position ids~\cite{vaswani2017attention}) maintaining operations and CPU-GPU interactions.

To tackle this issue, \sys incorporates \textit{masking} technique~\cite{vaswani2017attention}.
This technique is employed to ensure that the predictions for position $i$ depends only on the known outputs at positions less than $i$ in decoder-based LLMs.
The masking technique relies on a table where all the positions with value of one will be taken into account for calculation.

Crucially, the tree decoder retains the autoregressive procedure while only modifying its masking table to support the isolation of effects from different branches, as demonstrated in Figure~\ref{fig-tree-masking}(b).  
During each iteration, \sys treats the new generated token as input, just as in regular generation.
However, it assigns a value of one only to the previous positions on the same branch.
For example, when generating token ``7'' for branch b2 in iteration 5, the input remains $T_6$, the output of iteration 4,  but only the positions ``1, 2, 3, and 5'' are set to one; all other positions are set to zero.
This approach ensures that tokens ``4 and 6'' do not affect the calculation, enabling \sys to generate token ``7'' without being influenced.



\noindent \textbf{Token tree verification and rectification}
To achieve the goal of not sacrificing accuracy, \sys has to verify every token generated by the \smallmodel LLM.
An intuitive approach is to use the \largemodel LLM to run verification after each token generation by the \smallmodel LLM.
However, such an approach (called \textit{autoregressive verification (AV)}) is even slower than  generating every token by the \largemodel model directly, since AV does not reduce the \largemodel LLM inference times but even uses \smallmodel LLM.

\begin{figure}
    \centering
    \includegraphics[width=0.48\textwidth]{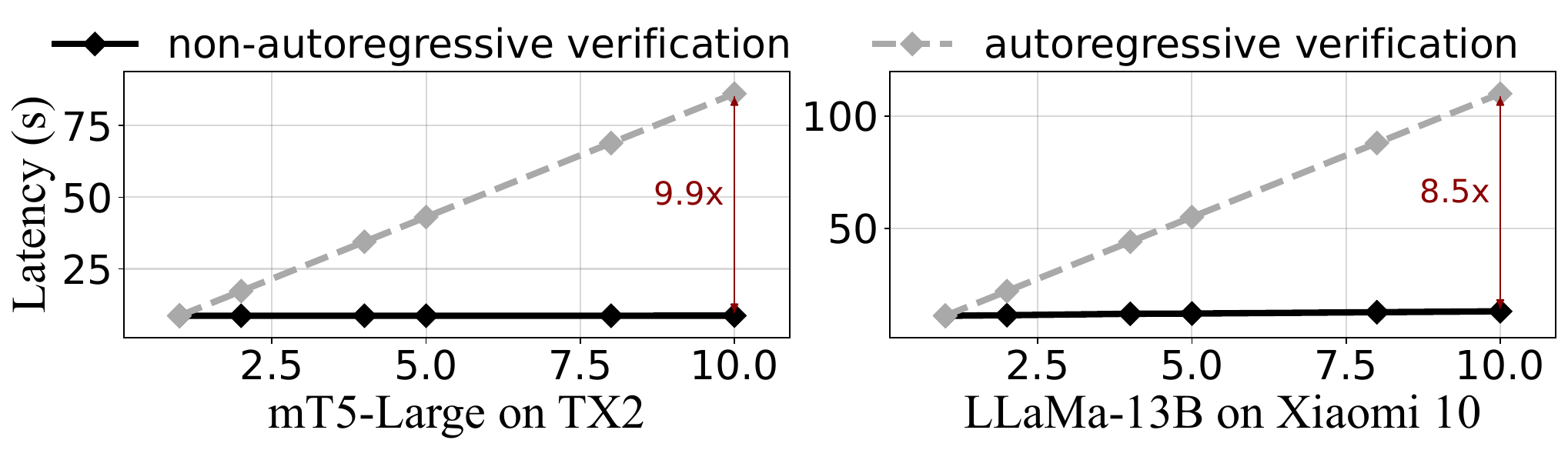}
    \vspace{-20pt}
    \caption{The latency of verification with input sequencing increasing. }
    \vspace{-20pt}
    \label{fig-non-autoregressive-verification}
\end{figure}





To tackle this issue, \sys is based on two opportunities:
(1) The \largemodel LLM can examine a sequence of tokens in parallel by visiting its parameters only once (called \textit{non-autoregressive verification (NAV)}) and the verification results are the same as examining them sequentially~\cite{chen2023accelerating,leviathan2023fast}.
(2) The NAV process is way faster than AV.
Our pilot experiments in Figure~\ref{fig-non-autoregressive-verification} on LLaMa-13B and mT5-Large models using Xiaomi10 and TX2 shows that NAV outperforms AV in examining time across 2--10 input tokens, with its benefits more pronounced as token count increases. 
NAV significantly reduces verification time for mT5-Large and LLaMa-13B models by 8.5–9.9$\times$ at 10 tokens, attributed to NAV's single weight swapping versus AV's multiple swappings per token, reducing I/O overhead.

To sum up, NAV can parallel verifying multiple tokens correctly at the cost of only one \largemodel LLM inference. 
\sys incorporates NAV and extends it to support token tree verification, as shown in Figure~\ref{fig-post-processing}, including two crucial steps:

\begin{itemize} [leftmargin=*,topsep=1pt]
    \item \textit{Batched non-autoregressive verification.}
    As shown in Figure~\ref{fig-post-processing}\textcircled{1}, to support token tree verification, \sys first divides the token tree into several branch sequences and combines them to a mini-batch as the input of the \largemodel LLM.
    After NAV, \sys can obtain the correct results of every position in each branch sequence.
    Compared with the origin branch sequences, \sys can detect all errors of a token tree, e.g., $T'_7$ in branch2 results.
    A branch correct sequence is the sub-sequence leading up to the first error position, plus the rectified token, to avoid error propagation.
    For example, branch1 correct sequence stops at the $T_2$, plus the $T_5$, i.e., tokens ``1, 2 and 5''.
    To be noted, given that the NAV is I/O-bounded, increasing batch size (e.g., <10) has a negligible effects on verification time. 
    \item \textit{Depth-first verified sequence searching.}
    Based on the correct sequences, \sys can build a correct token tree, as shown in Figure~\ref{fig-post-processing}\textcircled{2}.
    Its leaf node is either first rectified token or the origin branch sequence last token.
    \sys leverages depth-first searching algorithm to find the longest correct path in the correct token tree as the \textit{verified sequence}.
\end{itemize}

If the verified sequence has a rectified token, e.g., $T'_7$, \sys will rollback the token tree to the error position, fix the error, and use it as the new input for future generation.


\subsection{Self-adaptive fallback strategy} \label{sec-design-fallback-strategy} 
This strategy is devised to initiate the verification process promptly when the memory-resident LLM generates an incorrect token.
To achieve this goal, \sys needs to answer two crucial questions:
\begin{itemize} [leftmargin=*,topsep=0pt]
    \item \textbf{Selection of Decision Metric.}
    The decision metric should effectively evaluate the probability of errors within the token tree. 
    
    \item \textbf{Threshold Values for Different Tasks.} 
    Recognizing that a universal threshold may not be suitable for all tasks, \sys must establish appropriate threshold values tailored to specific tasks.

\end{itemize}

To tackle these issues, \sys introduces two innovative techniques:

$\bullet$ \textit{Tree-cumulative confidence ($T_c$).} 
We propose using \textit{tree-cumulative confidence} as the decision variable for initiating fallback.
Unlike prior studies~\cite{kim2023big,leviathan2023fast} that rely on a single token confidence or token sequence length, $T_c$ provides a comprehensive assessment of global uncertainty. 
It captures errors more accurately due to the autoregressive nature of token generation.

The formulation \textit{tree-cumulative confidence} is as 
$ T_c = max_{i=1}^{N_c} C_i$,
where $N_c$ represents the number of branches in a token tree, and $C_i$ denotes the cumulative confidence of the $i$-th branch.
We select the maximum cumulative confidence over minimum/average confidence because the most confident branch is more likely to yield the correct result after verification, and the verification process can only identity errors  when the most confident branch is wrong.

$\bullet$ \textit{self-adaptive threshold ($\alpha$)} is utilized to determine when \largemodel LLM shall verify.
It operates on the principle that the \smallmodel LLM, which generates outputs closely resembling those of the \largemodel LLM, should be trusted more, i.e., a lower verification frequency by setting  a lower threshold.
To assess the outputs similarity, \sys relies on historical data regarding the accuracy of verified tokens. 

Users can either select an initial $\alpha$ value or utilize the default value (0.01) provided by the system.
After verification, \sys updates the \textit{self-adaptive threshold ($\alpha$)} using the following rule:
\begin{align}
    \alpha_{i+1} = \left\{\begin{matrix}
        &\alpha_i * 0.5   &if\: N_{correct}  == N_{all} \\
        &\alpha_i / T_c^{\frac{N_{all}-N_{correct}}{N_{all}}}  &if\: N_{correct}  < N_{all}
    \end{matrix}\right.
\end{align}
where $N_{correct}$ and $N_{all}$ are the number of total tokens and correct tokens in the most matching branch during one verification.
Specifically, when the verification process detects no error, \sys lowers $\alpha$ by multiplying the current value by 0.5, the cumulative confidence of 3-5 tokens in empirical observations.
In contrast, if verification identifies errors, the threshold is increased by dividing $\alpha$ by the average cumulative confidence of all tokens subsequent to the incorrectly generated one.
The rationale behind the use of an exponential function is that the \textit{tree-cumulative confidence} is the product of every token's confidence, accumulating exponentially. 

In summary, after each token generation by the \smallmodel LLM, \sys calculates $T_c$.
If $T_c$ falls below $\alpha$, a fallback occurs, and the \largemodel model begins verification.
After verification, $\alpha$ is updated based on the latest generation accuracy history.

\subsection{Speculative Generation Pipeline} \label{sec-design-speculative}


As elaborated in $\S$\ref{sec-bgd-chanllenge}, the GPU utilization undergoes cyclical upswings attributed to the fact that SOTA LLM engines resort to the \textit{swapping} technique.
To harvest the free cycles, \sys proposes \textit{speculative generation} technique by allowing the \smallmodel LLM to continue generating tokens during verification process.
This approach is based on the insight that sometimes the verification detects no error so the speculatively generated tokens could be used afterwards.


\noindent \textbf{The impacts of speculative generation on \largemodel LLM.}
Our preliminary experiments on TX2 using mT5-Large and GPT2-Large models shows that paralleling the \smallmodel and \largemodel LLM execution increases the the \largemodel LLM computing and loading time by 2.2--2.3$\times$ and 1.05--1.09$\times$, respectively.
The computing delay is attributed to the GPU cores contention, while the loading delay is unexpected which is because of the memory contention.
Specifically, the \smallmodel LLM continually allocates memory regions for speculatively generated tokens, while the \largemodel LLM dynamically allocates memory regions for loading parameters.
Typically, negligible effects are exerted on each other unless the memory usage is over 90\%. 
However, it is common for memory usage to exceed 90\% or even reach 95\% in the speculative generation scenario, attributed to the fact that existing state-of-the-art LLMs engines are designed for loading as many parameters as possible from disk to the memory to reduce inference time.


\begin{figure}[t]
	\centering
	\includegraphics[width=0.48\textwidth]{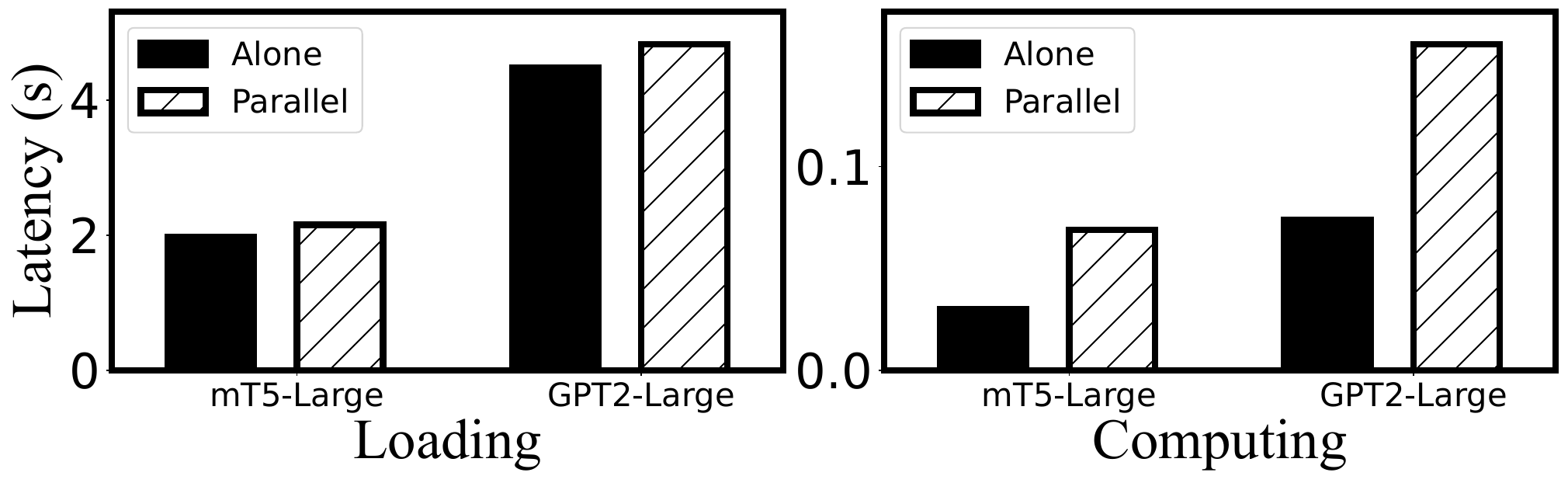}
    \vspace{-10pt}
	\caption{The loading and computing time of the \largemodel model execution with the \smallmodel model parallelization.}
    \vspace{-10pt}
	\label{fig-parallel-gpu}
\end{figure}
\begin{figure}
    \centering
    \includegraphics[width=0.96\linewidth]{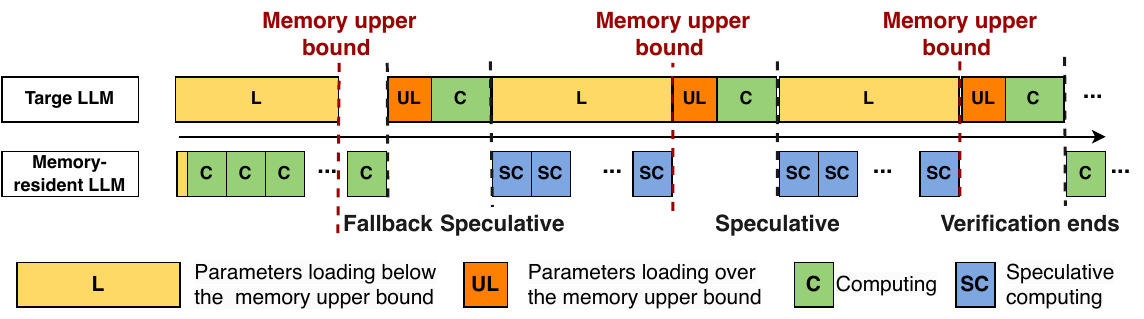}
    \vspace{-10pt}
    \caption{\sys's speculative generation.  
    }
    \vspace{-10pt}
    \label{fig-speculative-plan}
\end{figure}




\noindent \textbf{Computing-loading pipeline.} 
To tackle the above issue, \sys delicately plans the parallel execution, as shown in Figure~\ref{fig-speculative-plan}.
The main principle is that normal verification process cannot be influenced by the speculative execution.
Thus, there is a memory upper bound by profiling or user defining to avoid two LLMs memory contention, and the two LLMs computing cannot be executed in parallel.

After feeding the input sequence, the parameters of both the \smallmodel and the \largemodel LLMs are loaded to the memory.
Once the \smallmodel LLM loading finishes, it begins to generate tokens, and the \largemodel LLM parameters loading (in yellow) will stop before the memory upper bound is exceeded to avoid influencing normal \smallmodel LLM generation. 
When the fallback condition is met, the rest of parameters for the \largemodel model (in orange) will be loaded into the memory, and then the computing of the \largemodel LLM begins.
The speculative execution  (in blue)  will not run unless the verification process is loading parameters below the memory budget (in yellow), avoiding  processors and memory contention.

\section{evaluation}

\subsection{Implementation and Setups} \label{sec-evaluation-setting}
We have fully implemented \sys with 4.5k SLoC (Python: 3,500 and C/C++: 1,000).
The prototype is a standalone framework supporting LLMs exported from TensorFlow~\cite{tensorflow} and PyTorch~\cite{pytorch}.
\sys leverages llama.cpp~\cite{llama-cpp} (one of the most lightweight on-device LLM engine) as the smartphone backend and PyTorch~\cite{pytorch} as the IoT device backend.



\begin{table}[t]
\centering
\smaller
\resizebox{!}{1.5cm}{%
\begin{tabular}{lllr}
\toprule[1.5pt]
\textbf{Platform}       & \textbf{Processor}                                                                                                      & \textbf{Software}                                                                       & \multicolumn{1}{r}{\textbf{Mem.}} \\ \hline
\begin{tabular}[l]{@{}l@{}}  Jetson TX2  \end{tabular}     & \begin{tabular}[l]{@{}l@{}} 4x Cortex-A57 \\ Maxwell 128 CUDA cores \end{tabular}                                                                                                    &   \begin{tabular}[l]{@{}c@{}} Torch-1.10 \\ Ubuntu 18.04 \end{tabular}                                                           & 8G                                  \\ \hline
\begin{tabular}[l]{@{}l@{}} Jetson \\ Orin NX \end{tabular} & \begin{tabular}[l]{@{}l@{}}Ampere 1024 CUDA cores\\ + 32 Tensor Cores\end{tabular}                                                                                                   &  \begin{tabular}[l]{@{}c@{}} Torch-2.0 \\ Ubuntu 18.04 \end{tabular} & 8G                                 \\ \hline
Xiaomi 10     & \begin{tabular}[l]{@{}l@{}}1x 2.84GHz A77+3x 2.4GHz Cortex A77 \\ +4x 1.8GHz Cortex A55\end{tabular}           &          \begin{tabular}[l]{@{}c@{}} Android 10 \\ llama.cpp \end{tabular}                                                                 & 8G                                  \\ \hline
\begin{tabular}[l]{@{}l@{}} Xiaomi \\ 11  \end{tabular}    & \begin{tabular}[l]{@{}l@{}}1x 3.0 GHz X2+ 3x 2.5 GHz Cortex A710\\ + 1.8GHz 4x Cortex-A510\end{tabular} &              \begin{tabular}[l]{@{}l@{}} Android 10 \\ llama.cpp \end{tabular}                                & 8G                                  \\ \toprule[1.5pt]
\end{tabular}%
}
\caption{Platforms used in the experiments.}
\vspace{-30pt}
\label{table-platforms}
\end{table}
\begin{table}[t]
\centering
\resizebox{0.48\textwidth}{!}{%
\begin{tabular}{llllrl}
\toprule[1.5pt]
\textbf{Devices}                                                                           & \textbf{Tasks}      & \textbf{\begin{tabular}[c]{@{}l@{}}Memory-\\ resident LLMs\end{tabular}}          & \textbf{\begin{tabular}[c]{@{}l@{}}Target \\ LLMs\end{tabular}}                    & \textbf{\begin{tabular}[c]{@{}r@{}}Speed\\ Gap\end{tabular}} & \textbf{Datasets}                                                          \\ \hline
\multirow{6}{*}{\begin{tabular}[c]{@{}l@{}}Jetson TX2\\ \\ Jetson \\ Orin NX\end{tabular}} & \multirow{2}{*}{T}  & mT5-small  (0.3B)                                                                 & mT5-Large  (1.2B)                                                                  & 230x                                                         & IWLST17-de-en~\cite{cettolo-etal-2017-overview}                                                              \\
                                                                                           &                     & Bart-base                                                                         & Bart-Large                                                                         &                                                              & WMT14-de-en~\cite{bojar-EtAl:2014:W14-33}                                                                \\ \cline{2-6} 
                                                                                           & \multirow{2}{*}{QA} & mT5-small  (0.3B)                                                                 & mT5-Large  (1.2B)                                                                  & 230x                                                         & SQuAD\_v2~\cite{2016arXiv160605250R}                                                                  \\
                                                                                           &                     & T5-small (0.06B)                                                                  & T5-large  (0.73B)                                                                  & 263x                                                         & SQuAD\_v2                                                                  \\ \cline{2-6} 
                                                                                           & LM                  & GPT2  (0.14B)                                                                     & GPT2-Large  (0.8)                                                                  & 214x                                                         & Wikitext~\cite{merity2016pointer}                                                                   \\ \cline{2-6} 
                                                                                           & S                   & T5-small (0.06B)                                                                  & T5-large (0.73B)                                                                   & 263x                                                         & CNN/Daily~\cite{see-etal-2017-get}                                                                  \\ \toprule[1.5pt]
\multirow{3}{*}{\begin{tabular}[c]{@{}l@{}}Xiaomi 10\\ \\ Xiaomi \\ Pro\end{tabular}}      & T                   & \begin{tabular}[c]{@{}l@{}}Vicuna-7B\\ (INT4)\end{tabular}                        & \begin{tabular}[c]{@{}l@{}}Vicuna-13B\\ (INT4)\end{tabular}                        & 59x                                                          & \begin{tabular}[c]{@{}l@{}}Parrot\\ WMT22-de-en\\ WMT22-zh-en\end{tabular} \\ \cline{2-6} 
                                                                                           & QA                  & \multirow{2}{*}{\begin{tabular}[c]{@{}l@{}}LLaMa2-Chat-\\ 7B (INT4)\end{tabular}} & \multirow{2}{*}{\begin{tabular}[c]{@{}l@{}}LLaMa2-Chat-\\ 13B (INT4)\end{tabular}} & 59x                                                          & \begin{tabular}[c]{@{}l@{}}SQuAD\\ TruthfulQA~\cite{lin2021truthfulqa}\end{tabular}                 \\ \cline{2-2} \cline{6-6} 
                                                                                           & S                   &                                                                                   &                                                                                    & \multicolumn{1}{l}{}                                         & CNN/Daily                                                                  \\ \toprule[1.5pt]
\multicolumn{6}{l}{ \begin{tabular}[l]{@{}l@{}}T, S, QA, and LM represent the generative tasks of translation, summary, \\ question-answering, and language modeling. \end{tabular}}                                                                                                                                                                                                                                                                                              
\end{tabular}%
}
\caption{Tasks, models, datasets, and their corresponding tested devices used in the experiments.}
\vspace{-20pt}
\label{tab-model-datasets-tasks-devices}
\end{table}

\noindent \textbf{Hardware setup.}
We test the performance of \sys on four devices: 2 smartphones (Xiaomi 10 and Xiaomi 12) and 2 IoT devices (Jetson TX2, and Jetson Orin), as summarized in Table~\ref{table-platforms}.
We run LLMs on Jetson GPUs and smartphone CPUs, since the existing LLM engines [refs] have unmature support for smartphone GPU/NPU.
Nevertheless, \sys's design is orthogonal to hardware types.

\noindent \textbf{Models and datasets.}
We test with a range of typical LLM models with various generative tasks datasets across different devices, as summarized in Table~\ref{tab-model-datasets-tasks-devices}. 
On the IoT devices, we evalaute \sys on two translation tasks, two question answering tasks, one language modeling task, and one summarization tasks with mT5, T5, GPT2, and Bart models.
All the models  are fine-tuned by ourselves as ~\cite{kim2023big} does.
For smartphone devices, we use Vicuna-1.5 and LLaMA2 models with three translation tasks, one question answering task and , and one summarization.
All the models are downloaded from hugging face repository~\cite{TheBloke} and have been quantized by AutoGPTQ~\cite{AutoGPTQ} into 4-bit format for saving memory and improving inference speed.

\begin{table*}[t]
\centering
\begin{subtable}[t]{0.48\textwidth}
\resizebox{!}{1.5cm}{
\begin{tabular}{llrrrr}
\toprule[1.5pt]
\textbf{Models}                             & \textbf{Datasets}             & \textbf{StdPL}  & \textbf{SP} & \textbf{BLD} & \textbf{Ours} \\ \toprule[1.5pt]
\multirow{2}{*}{mT5-Large} & T: IWLST17-de-en & 100      & 100 & 96.1         & 100           \\
                                    & QA: SQuAD        & 100       & 100 & 52.9            & 100           \\ \hline
\multirow{2}{*}{T5-Large}  & S: CNN/Daily     & 100       & 100  &     5.5     & 100           \\
                                    & QA: SQuAD        & 100      & 100  & 51.4           & 100           \\ \hline
Bart-Large               & T: WMT14-de-en   & 100        & 100 & 96.5         & 100           \\ \hline
GPT2-Large                 & LM: Wikitext     & 100      & 100  &    12.9      & 100           \\ \toprule[1.5pt]
\end{tabular}%
}
\caption{IoT devices}
\end{subtable}
\begin{subtable}[t]{0.49\textwidth}
\resizebox{!}{1.5cm}{
\begin{tabular}{llrrrrr}
\toprule[1.5pt]
\textbf{Models}                             & \textbf{Datasets} & \textbf{StdPL} & \textbf{STI} & \textbf{SP} & \textbf{BLD} & \textbf{Ours} \\ \toprule[1.5pt]
\multirow{3}{*}{ \begin{tabular}[l]{@{}l@{}} Vicuna-13B \\ (INT4) \end{tabular}}  & T: Parrot        & 100     & 86.7 & 100   & 89.7         & 100           \\
                                           & T: WMT22-de-en   & 100     &  87.2  & 100   & 90.2         & 100           \\
                                           & T: WMT22-zh-en   & 100      & 88.1  & 100   & 80.4         & 100           \\ \hline
\multirow{3}{*}{ \begin{tabular}[l]{@{}l@{}} LLaMa2-Chat \\ (INT4) \end{tabular} } & S: CNN/Daily     & 100          &  81.0  & 100        & 7.96    & 100       \\
                                           & QA: SQuAD        & 100       & 83.2  & 100  & 3.4          & 100           \\
                                           & QA: Truthful\_QA & 100       & 85.4  & 100  & 4.7          & 100           \\ \toprule[1.5pt]
\end{tabular}%
}
\caption{Smartphones}
\end{subtable}
\vspace{-10pt}
\caption{The summary of the generation accuracy of \sys and the baseline on tested devices. T:*, S:*, QA:*, and LM:* represent the generative tasks of translation, summary, question-answering, and language modeling.}
\vspace{-20pt}
\label{table-accuracy-tasks-smartphones}
\end{table*}
\begin{figure*}
    \centering
    \begin{subfigure}[t]{0.96\textwidth}
        \includegraphics[width=\textwidth]{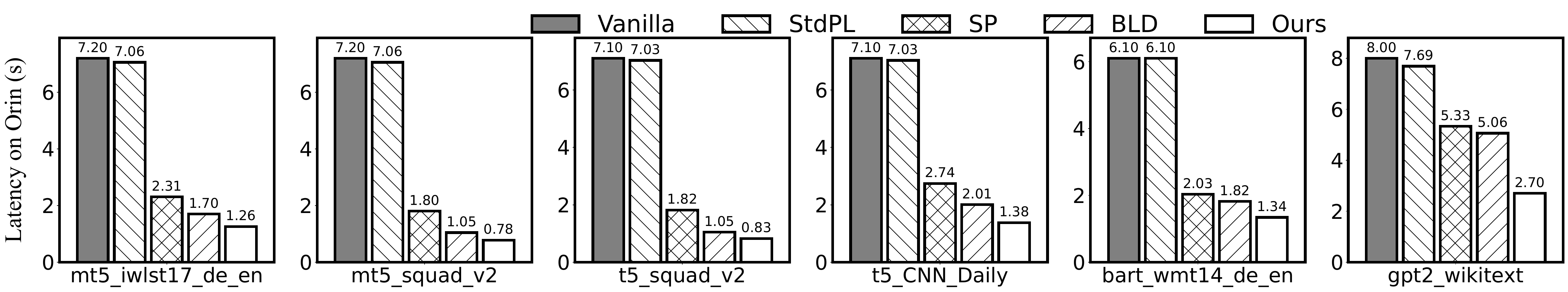}
    \end{subfigure}
    \begin{subfigure}[t]{0.96\textwidth}
        \includegraphics[width=\textwidth]{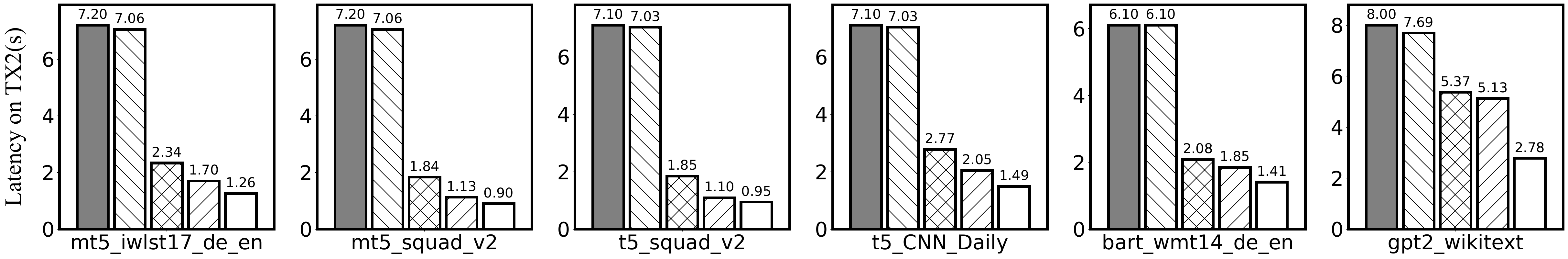}
    \end{subfigure}
    \vspace{-10pt}
    \caption{Average per-token generation latency of \sys and the baselines under different tasks on IoT devices.}
    \vspace{-10pt}
    \label{fig-model-inference-tx2}
\end{figure*}


\noindent \textbf{Baselines.}
We mainly compare \sys with 5 state-of-the-art baselines which can be divided into two categories:

\noindent $\bullet$ 3x \textit{Single-LLM baselines. }
(1) \texttt{Standard (Std)} always utilizes the \largemodel LLM to generate outputs, with PyTorch for IoT and llama.cpp for smartphones.
(2) \texttt{Standard pipeline (StdPL)}: It executes a layer-wise pipeline, overlapping I/O and computation, as used by existing SOTA LLM inference engines.
(3) \texttt{Speedy Transformer Inference (STI)~\cite{guo2023sti}}: An edge NLP inference framework with quantization parameter shards and fine-grained computing-loading pipeline.

\noindent $\bullet$ 2x \textit{LLM collaboration baselines. }
(1) \texttt{Speculative Decoding (SP)}~\cite{kim2023big}: A state-of-the-art framework also uses ``generator and verifier'' LLM collaboration. 
(2) \texttt{Big Little Transformer Decoder (BLD)}~\cite{kim2023big}: An algorithm of determining verification timing and rollback mechanism for ``generator and verifier'' LLM collaboration. 


\noindent \textbf{Metrics and configurations.}
We mainly report generation accuracy and the per-token generation time.
For clarity, \sys's goal is to align the \smallmodel LLM outputs to the \largemodel LLM.
Thus, we regard the text produced by the \largemodel LLM as the ground truth and calculate the Rouge-L score~\cite{lin2004rouge}, a similarity between two sequences based on the longest common subsequence, as the \textit{generation accuracy}.



\subsection{Generation Speed}
\begin{figure*}
    \centering
    \begin{subfigure}[t]{0.96\textwidth}
        \includegraphics[width=\textwidth]{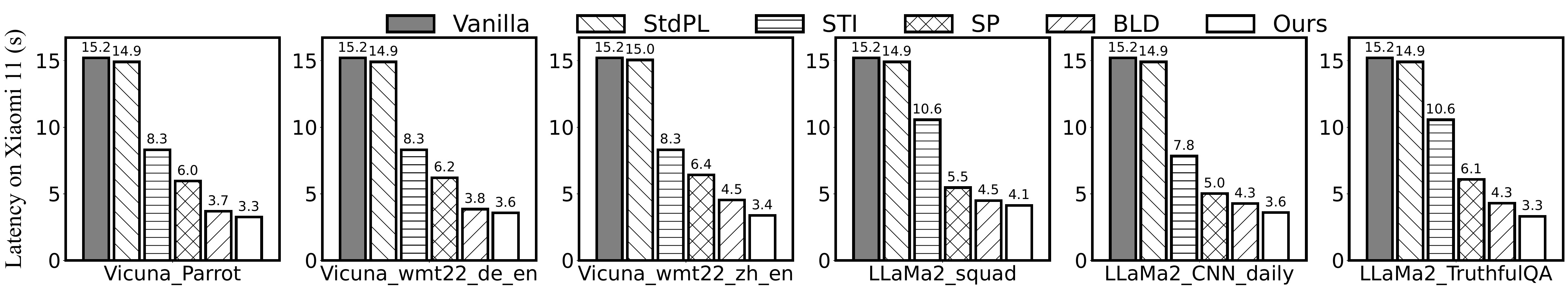}
    \end{subfigure}
   \begin{subfigure}[t]{0.96\textwidth}
       \includegraphics[width=\textwidth]{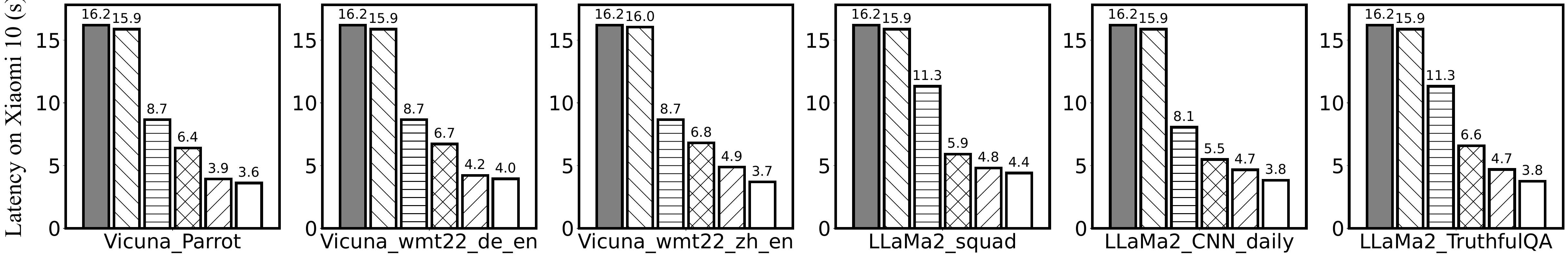}
   \end{subfigure}
    \vspace{-10pt}
    \caption{Average per-token inference latency of \sys and the baselines under different tasks on smartphones.}
    \vspace{-5pt}
    \label{fig-model-inference-xiaomi}
\end{figure*}

\noindent \textbf{Overall performance.}
We first comprehensively investigate the generation performance of \sys on four tested devices.
The generation accuracy and per-token generation time results are illustrated in Table~\ref{table-accuracy-tasks-smartphones}, Figure~\ref{fig-model-inference-tx2} and Figure~\ref{fig-model-inference-xiaomi}, respectively. 
Our key observation is that \textbf{\sys consistently and remarkably outperforms other baselines on per-token generation time without comprising accuracy across all tested devices.}

$\bullet$ \textit{Generation time of \sys v.s. Single-LLM baselines.}
Compared with Std, \sys achieves a 2.9-9.3$\times$ and 3.47--4.67$\times$ speedup in per-token average generation time on IoT and smartphone devices, respectively, without compromising accuracy.
Specifically, \sys can generating question-answering outcomes on Xiaomi 11 at a fastest speed of 0.86 s/token.
This achievement enables real-time token generation with over-10B LLM on COTS device for the first time.
This is attributed to the fact that \sys can delegate most of token generations to the \smallmodel LLM and ensure correctness by non-autoregressive verification. 

When compared with more competitive baselines like StdPL and STI, \sys reduce per-token average generation time 2.9-9.3$\times$ and  1.83--2.45$\times$, respectively.
Those benefits are attributed to the fact employing \smallmodel LLMs for text generation consistently outpaces any pipeline or quantization approaches of \largemodel LLMs on the mobile device, where \smallmodel LLM can yield a over hundredfold speed improvement compared to the \largemodel LLM.
Besides, it can also improve generation accuracy by 11.1--19.0 percentage point, compared to STI.
This benefits from our tree non-autoregressive verification which can examine and correct all errors by the \smallmodel LLM efficiently.

$\bullet$ \textit{Generation time of \sys v.s. LLM collaboration baselines.}
Compared with BLD, \sys can achieves a 4.5--94.5 and 9.8--96.7 percentage point generation accuracy improvement with a 1.1-1.4$\times$ and 1.1--1.3$\times$ speedup in per-token average generation time on IoT and smartphone devices, respectively.
That is because, unlike BLD which speeds up the generation process by reducing the number of correction (sacrificing accuracy), our self-adaptive fallback strategy aims to minimize verification times while ensuring verification for each token. 
Such an approach enhances generation speed without sacrificing accuracy.
Furthermore, speculative execution enables \smallmodel LLM to generate texts earlier without waiting the verification results when no errors are detected by the verification process, further reducing generation latency.

Similarly, \sys can reduce  per-token average generation time by 1.93--2.00$\times$ and 1.34--1.77$\times$ on IoT and smartphone devices, respectively.
That is because, unlike SP uses token sequence length, our self-adaptive fallback strategy can accurate finds when the \smallmodel LLM generate errors and can reduce verification frequency.

\begin{figure}
    \centering

    \begin{subfigure}[t]{0.48\textwidth}
        \includegraphics[width=\textwidth]{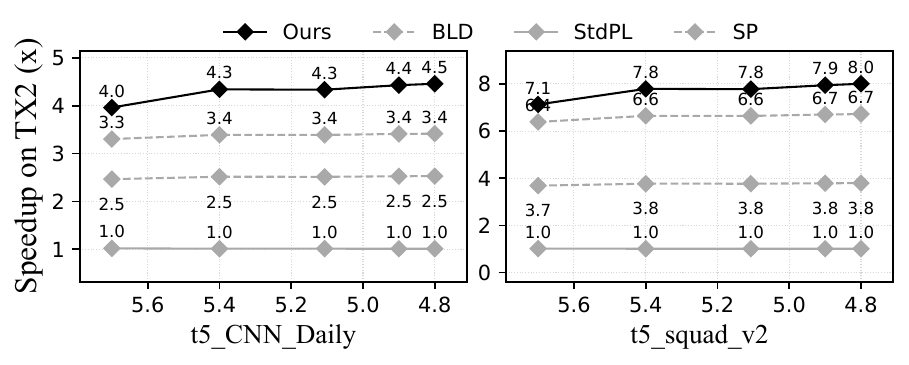}
    \end{subfigure}
    \begin{subfigure}[t]{0.48\textwidth}
        \includegraphics[width=\textwidth]{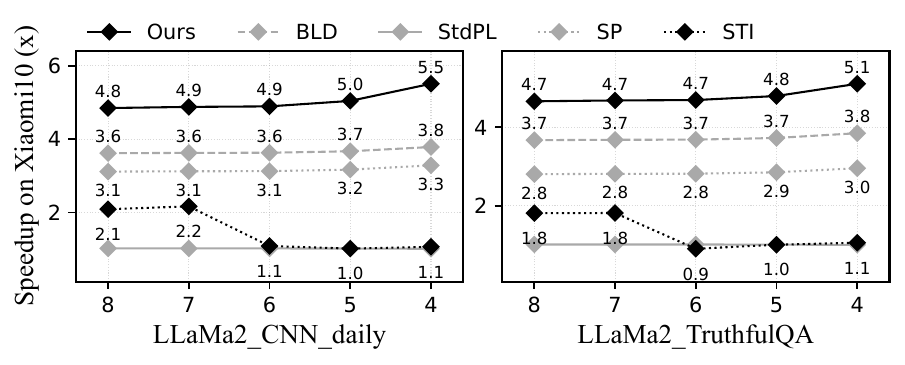}
    \end{subfigure}
    \vspace{-20pt}
    \caption{The speedup of different baselines under different memory budgets.}
    \vspace{-20pt}
    \label{fig-memory-budget}
\end{figure}
\begin{table*}[t]
\centering
\begin{subtable}[t]{0.48\textwidth}
\resizebox{!}{1.4cm}{%
\begin{tabular}{lrrrrr}
\toprule[1.5pt]
\textbf{\begin{tabular}[c]{@{}l@{}}Models-tasks-datasets\end{tabular}}             & \textbf{Vanilla} & \textbf{StdPL}  & \textbf{SP}   & \textbf{BLD} & \textbf{Ours}        \\ \toprule[1.5pt]
\begin{tabular}[c]{@{}l@{}}mT5-translation\\ IWLST17-DE-\textgreater{}EN\end{tabular} & 36.9             & 36.2                    & 12.0 & 7.7          & \textbf{7.7 (4.8$\times$)}  \\ \hline
\begin{tabular}[c]{@{}l@{}}T5-summary\\ CNN/Daily\end{tabular}                        & 36.4             & 36.0                   & 7.6  & 10.3         & \textbf{8.4 (4.3$\times$)}  \\ \hline
\begin{tabular}[c]{@{}l@{}}T5-QA\\ SQuAD\end{tabular}                                 & 36.9             & 36.5                    & 15.4 & 9.9          & \textbf{4.6 (8.0$\times$)}  \\ \toprule[1.5pt]
\end{tabular}%
}
\caption{Jetson Orin NX}
\end{subtable}
\begin{subtable}[t]{0.48\textwidth}
\resizebox{!}{1.4cm}{%
\begin{tabular}{lrrrrrr}
\toprule[1.5pt]
\textbf{\begin{tabular}[c]{@{}l@{}}Models-tasks-datasets\end{tabular}}             & \textbf{Vanilla} & \textbf{StdPL} & \textbf{STI} & \textbf{SP}   & \textbf{BLD} & \textbf{Ours}        \\ \toprule[1.5pt]
\begin{tabular}[c]{@{}l@{}}LLaMa2-summarization\\ CNN/Daily mail\end{tabular}         & 56.2             & 55.1           & 27.9         & 21.5 & 18.6         & \textbf{17.3 (3.2$\times$)} \\ \hline
\begin{tabular}[c]{@{}l@{}}LLaMa2-QA\\ TruthfulQA\end{tabular}                        & 56.2             & 55.1           & 28.1         & 23.9 & 18.3         & \textbf{14.3 (3.9$\times$)} \\ \hline
\begin{tabular}[c]{@{}l@{}}Vicuna-translation\\ WMT22-DE-EN\end{tabular}              & 56.2             & 55.1           & 20.7         & 20.4 & 20.3         & \textbf{15.5 (3.6$\times$)} \\ \toprule[1.5pt]
\end{tabular}%
}    
\caption{Xiaomi 11}
\end{subtable}
\vspace{-5pt}
\caption{The summary of the energy consumption (J) of different models across different devices.}
\vspace{-20pt}
\label{table-energy-devices}
\end{table*}

\subsection{Memory Sensitivity Analysis}
This subsection is to investigate the impact of different memory budgets on our approach.
We further conduct experiments on mT5 and T5 models on TX2 and LLaMa2 on Xiaomi 10 respectively under different memory budgets (e.g., from 4GB to 8GB on Xiaomi 10).
The speedup of different baselines are shown in Figure~\ref{fig-memory-budget}.
\textbf{\sys consistently exhibits the highest speedup among all baselines from 8GB to 4GB and its benefits are more prominent with the decreasing memory budget.}

\sys reduces generation time under 6GB memory budget on Jetson TX2 by 5.54$\times$, 1.76$\times$ and 1.12$\times$ on average for \texttt{StdPL}, \texttt{SP} and \texttt{BLD}, respectively; while the speedups for 4GB memory budget are 6.12$\times$, 1.91$\times$ and 1.25$\times$, correspondingly, which are 1.29$\times$, 1.08$\times$ and 2.1$\times$ larger than that under 6GB on average.
Similarly, \sys achieve a generation time speedup under 4GB memory budget on xiaomi 10 by 4.72$\times$ and 1.34$\times$ on average for \texttt{StdPL} and \texttt{BLD}, respectively.
That is because when the memory budget is stricter, the inference speed gap between the \smallmodel and \largemodel LLM is more significant, and delegating tokens to the \smallmodel LLM can gain more benefits.

\subsection{Energy Consumption Analysis}
We then evaluate the energy consumption of \sys with mT5 and T5 models on IoT devices and Vicuana and LLaMa2 model on smartphones.
As shown in Table~\ref{table-energy-devices}, compared with Std, StdPL, SP and BLD, \sys reduces per-token energy consumption by 4.35--7.96, 4.34--7.92, 1.56--3.33, and 1.05--2.15$\times$ on Jetson Orin NX, \sys achieves a energy consumption reduction by 3.22--3.59, 3.18--3.56, 1.24--1.66, 1.07--1.31 and 2.01--2.56,$\times$ correspondingly on Xiaomi 11, plus STI.
This is because \sys's two techniques can delegate as many tokens as possible to \smallmodel LLM while not sacrificing accuracy.

Compared with the latency speedup, \sys's energy consumption is relatively lower.
This situation arises because our speculative generation parallels the \smallmodel and \largemodel LLM execution together, resulting in more energy consumption.

\begin{figure}
    \centering
    \includegraphics[width=0.48\textwidth]{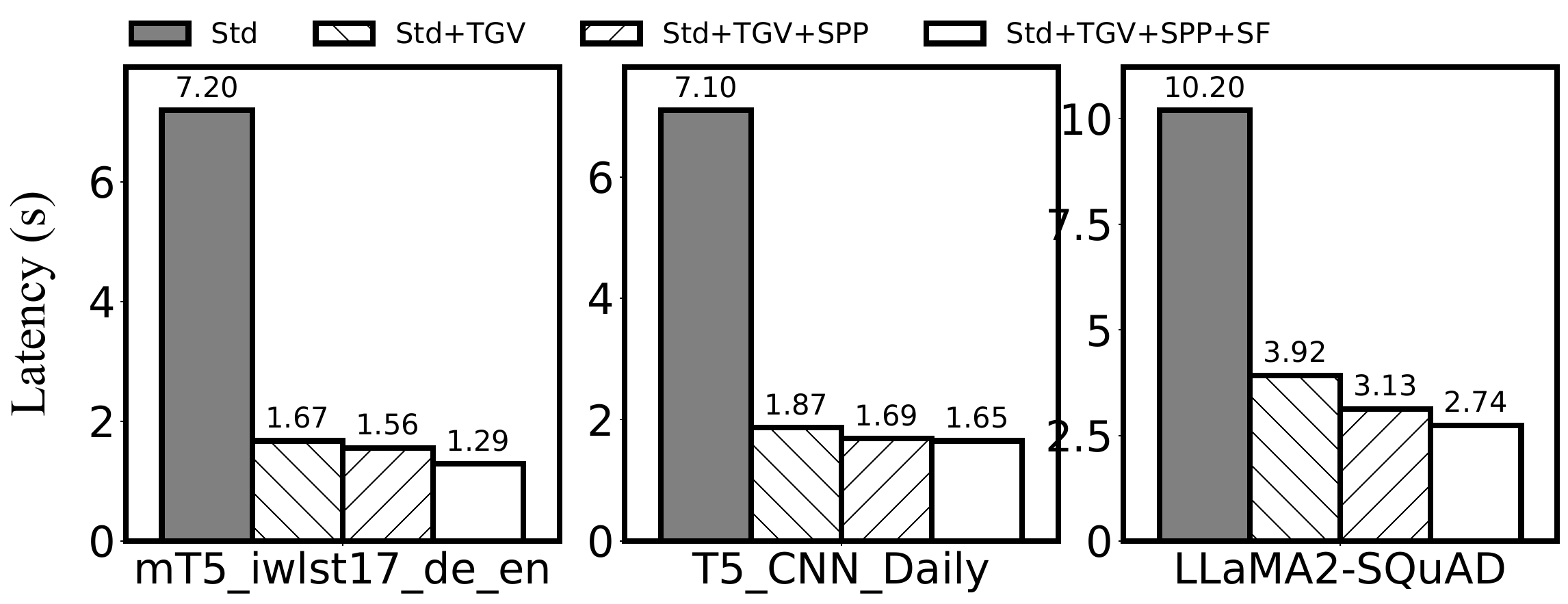}
    \vspace{-20pt}
    \caption{Ablation study of \sys.}
    \vspace{-20pt}
    \label{figure-ablation}
\end{figure}
\subsection{Ablation Study}
\noindent \textbf{Overall techniques.} We further conduct a breakdown analysis of the benefit brought by each of \sys’s techniques. 
The experiments are performed with the mT5 and T5 models on TX2 and LLaMa2 on Xiaomi 10.
The results are illustrated in Figure~\ref{figure-ablation}.
The leftmost bar is the same as baseline \texttt{Vanilla}, while the leftmost one is \sys.
The three crucial technique \textit{token tree generation and verification} in $\S$\ref{sec-design-tree-generation-verification}, \textit{self-adaptive fallback strategy}$\S$\ref{sec-design-fallback-strategy}  and the \textit{speculate generation} in $\S$\ref{sec-design-speculative} are represented by TGV, SF and SPP correspondingly.


We observe that \textbf{all techniques make a non-trivial contribution to the improvement.}
First, \textit{tree non-autoregressive generationa and verification} can delegete most token generations to the \smallmodel LLM, leading to an 2.6--4.3 speedup for mT5, T5 and LLaMa2 models, respectively.
The more benefits for mT5 model on IWLST translation dataset are because the mT5-Small model can generate more correct tokens than other two models so that more tokens can be delegated to the \smallmodel LLM. 
Besides, \textit{speculative execution} can reduce per-token generation time by up to 1.51$\times$.
That is because \sys can directly use the speculative results when the verification detects no error, especially true for LLaMA2 model.
Lastly, \textit{Self-adaptive fallback strategy} achieves a  1.08--1.20$\times$ speedup.
This is achieved by leveraging tree-cumulative confidence to assess error probabilities and dynamically adjust verification timings in response to variations in task complexity.



\section{related work}
\noindent \textbf{Model collaboration}
is a common optimization technique utilized to reduce inference latency~\cite{wang2023tabi,lee2019mobisr}.
Its key idea is to delegate most of the workloads to lightweight models to reduce inference latency while maintaining relatively high accuracy.
Tabi~\cite{wang2023tabi} is a multi-level inference engine that serves queries using various small models according to the queries difficulties.
MobiSR~\cite{lee2019mobisr} and NestDNN~\cite{fang2018nestdnn} employ the similar idea but depending on either resolutions or available resources.
Other ``early exit'' works~\cite{schuster2022confident,tambe2021edgebert,xin2020deebert,zhou2020bert}, which propose adaptive  exit timing relying on input data difficulties can also be regarded as a collaboration.
However, they either focus on CNN/encoder-based model architectures or must modify and retrain the model, hardly fitting into on-device LLM inference.
The most closely related works are \textit{speculative decoding}~\cite{leviathan2023fast,kim2023big,chen2023accelerating,miao2023specinfer}, which also employs smaller LLMs for text generation and larger LLMs for text verification.
\sys is motivated by these works and is the first inference engine for on-device generative NLP tasks, considering mobile devices unique challenges like the memory-bound situation.

\noindent \textbf{Mobile ML optimization.}
Machine learning optimization approaches, such as \textit{model compression}~\cite{yao2022zeroquant,frantar2022gptq,guan2021cocopie,niu2020patdnn,wang2020minilm,huynh2017deepmon,you2022speechmoe2,pope2023efficiently}, reducing the model size by quantization and knowledge distillation, \textit{caching}~\cite{xu2018deepcache,wang2020convergence,zhang2021elf}, reducing computation by reusing existing results, and \textit{token pruning}~\cite{wang2021spatten,cai2022enable,kim2022learned,rao2021dynamicvit,bolya2022token}, reducing computation by pruning useless tokens, have been extensively researched to reduce the generation latency.
\sys is orthogonal to and compatible with those algorithm-level optimizations.

Besides, some of researchers focus on generate text in a non-autoregressive manner~\cite{kasai2020deep,kong2022multilingual}.
However, these works can only apply to <1B models and have accuracy degradation problems, not the mainstream research direction.

\noindent \textbf{Pipeline optimization for ML.}
Pipeline optimization has been extensively used to accelerate ML~\cite{guo2023sti,bai2020pipeswitch,narayanan2019pipedream,wang2022overlap,zhou2022pets}.
Most of them, such as PipeDream~\cite{narayanan2019pipedream}, are utilized to scale out ML to multiple machines by pipelining forward/backward computation with activation/gradients synchronization to minimize bubbles of I/O and network communication.
Still, there are some studies focusing on single machine/task optimization.
For instance, PipeSwitch~\cite{bai2020pipeswitch} introduce pipelining model transmission over the PCIe and task execution in the GPU to reduce context switching performance overhead; while STI~\cite{guo2023sti} pipelines model shards loading with its computation to reduce inference latency.
\sys is inspired by these efforts and propose an efficient \textit{speculative generation pipeline} to address the challenge of I/O blocking and limited parallelism.

\section{conclusions}
This work has proposed \sys, the first efficient inference engine for on-device generative NLP tasks.
It breaks the memory wall and deliver LLM’s scaling ability to mobile devices
It incorporates three novel techniques, including: token tree generation and verification, Self-adaptive fallback strategy and speculative generation pipeline that can exploit the waste hardware resources during the verification process.
Our experiments  have demonstrated that when compared to the state-of-the-art LLM engines, \sys can reduce average per-token generation time by 2.9--9.3$\times$ and 3.5--4.7$\times$ on IoT devices and smartphones, without comprising accuracy.

\bibliographystyle{plain}
\bibliography{acmart}
\end{document}